\documentclass
[aps,preprint,onecolumn,showpacs,showkeys,10pt,letterpaper]{revtex4}%
\usepackage{amsfonts}
\usepackage{amsmath}
\usepackage{amssymb}
\usepackage{graphicx}
\usepackage{caption}%
\begin{document}
\title{Numerical approach to the nuclear deformation energy}
\author{B. Mohammed-Azizi}
\affiliation{University of Bechar, Bechar, Algeria}
\email{aziziyoucef@voila.fr}

\begin{abstract}
A numerical method close to the Strutinsky procedure (but better) is proposed
to calculate the deformation energy of nuclei. Quadrupole (triaxial)
deformations are considered. Theoretical as well as practical aspects of the
method are reviewed in this paper. A complete fortran program illustrates the
feasibility of the method. Thus, this code will constitute a useful "ready
tool" for those which deal with numerical methods in theoretical nuclear physics.

\end{abstract}
\volumeyear{year}
\volumenumber{number}
\issuenumber{number}
\eid{identifier}
\date[Date text]{date}
\received[Received text]{date}

\revised[Revised text]{date}

\accepted[Accepted text]{date}

\published[Published text]{date}

\startpage{1}
\endpage{ }
\maketitle
\volumeyear{ }

\section{Indroduction}

There are two methods which allow to determine the equilibrium shape ( ground
state) of the nuclei: The constrained Hartree-Fock method and the so-called
macroscopic-microscopic method. Though the latest generation of computers is
able to perform very complicated calculations, in terms of running time, it is
no so obvious to make systematic calculations for a large number of nuclei. A
good alternative is to use the Strutinsky method. The latter consists of
associating the classical liquid drop model with some shell and pairing
corrections built from a realistic microscopic model. Based on such a model,
we present a numerical method with its associated fortran program . The
potential energy of deformation is deduced as a function of the shape of the
nucleus. Triaxial (quadrupole) shapes are considered in this work. The three
semi axes of the ellipsoid are in fact connected to the both Bohr parameters
which are actually used in the calculations.

The different steps of calculations are:

i)The energy of deformation of the liquid drop model is first calculated [4].

ii)The Schrodinger equation of a microscopic Hamiltonian is built and solved
to obtain eigenvalues and eigenvectors. In fact we use the Fortran program
named "triaxial" already published in cpc. The microscopic model is explained
in details in this paper and also in Ref.[3].

iii) The semiclassical energy is deduced from the same Hamiltonian as (ii) is
calculated on the basis of the Wigner-Kirkwood expansion [5].

iv)The shell correction is deduced as the difference between the sum of
single-particle (point (ii)) energies and the same quantity smoothed
semiclassically (point (iii)).

\section{Potential energy of deformation in the liquid drop model}

We use the so-called macroscopic-microscopic method \cite{1967strutinsky} to
evaluate the potential energy of deformation of the nucleus. This method is
based on the liquid drop model plus shell and pairing corrections deduced from
a microscopic model \cite{2004azizi}.\newline The deformation (or potential)
energy of the nucleus is defined as the difference between the binding energy
of the deformed drop and the non-deformed drop (nucleus).
\begin{equation}
E(\beta)=E_{LD}(\beta)-E_{LD}(0) \label{209}%
\end{equation}
Here $\beta$ is a set of parameters defining the deformation. The case
$\beta=0$ represents the spherical shape (i.e., the non-deformed nucleus). We
recall that in the liquid drop model, the minimum is always obtained for the
spherical deformation. This involves $E(\beta)\geq0$. Of course, the liquid
drop or weizsaker formula model contains several terms, but only two depend on
the deformation of the nucleus, namely the surface and the coulomb energies.
Consequently, the other terms do not survive in the difference given by Eq.
(\ref{209}). The liquid drop energy reads \cite{1973pauli}%
\begin{equation}
E(\beta)=\frac{3}{5}\frac{e^{2}}{r_{0}}\frac{Z^{2}}{A^{1/3}}\left[  \frac
{A}{2Z^{2}}\zeta(B_{s}-1)+(B_{c}-1)\right]  \label{210}%
\end{equation}
with $r_{0}=1.275$ $fm$ and $e^{2}=1.4399764$ $MeV$. The quantities $B_{s}$,
et $B_{c}$ are the surface and the coulomb contributions. It is to be noted
that $B_{s}$ and $B_{c}$ are dimentionless and normalized to the unity so that
the deformation energy of the non deformed nucleus is equal to zero (i.e.,
$E(0)=0$). The reduced fissility has been determined empirically
\cite{1973pauli}:\newline$\zeta=52.8(1-2.84I^{2})$, $I=(N-Z)/(N+Z)$\newline
For triaxial ellipsoidal shape with semi axes $a,b,c$, the coulomb and surface
contributions are deduced analytically with the help of elliptic integrals of
the first and second kind $F(\varphi,k)$ et $E(\varphi,k)$ so that if
$a\geqslant b>c$, we will have \cite{1973pauli}:%
\begin{align}
B_{c}  &  =F(\varphi,k)a^{2}b^{2}c^{2}R_{0}^{-5}/(a^{2}-c^{2})^{1/2}%
,\label{211}\\
\sin\varphi &  =(1-c^{2}/a^{2})^{1/2},\ \ k^{2}=(a^{2}-b^{2})/(a^{2}%
-c^{2})\nonumber
\end{align}%
\begin{align}
B_{s}  &  =\frac{1}{2}R_{0}^{-2}\left\{  c^{2}+b(a^{2}-c^{2})^{-1/2}\left[
(a^{2}-c^{2})E(\varphi,k^{\prime})+c^{2}F(\varphi,k^{\prime})\right]  \right\}
\label{212}\\
\sin\varphi &  =(1-c^{2}/a^{2})^{1/2},\ \ k^{\prime2}=a^{2}b^{-2}(b^{2}%
-c^{2})/(a^{2}-c^{2})\nonumber
\end{align}
the condition of the volume conservation of the nucleus being $abc=R_{0}^{3}$
(equal to $r_{0}^{3}A$). with this condition it is clear that only two
deformation parameters are necessary to specify the shape of the nucleus. The
Bohr parameters $\left(  \beta,\gamma\right)  $ are more commonly employed in
this type of calculation. For moderate deformations, the link between the semi
axes and the Bohr parameters is given in Ref. \cite{2004azizi}. The elliptic
integrals are evaluated with Gauss quadrature formulae with 64 points.

\section{Shell correction}

According to the Strutinsky prescription, the shell correction to the liquid
drop model is defined as the difference between the sum of the single-particle
energies of the occupied states and the "smoothed part" of the same quantity:%

\[
\delta E=\sum_{occupied}\epsilon_{i}-\overline{\sum\epsilon_{i}}%
\]
In fact the Strutinsky procedure is done in such a way that the smoothed sum
does anymore contains shell effects so that the above difference represents
only the contribution due to the shell structure. In the Strutinsky's method
the smoothed sum is derived through the smoothed density of states
\cite{1967strutinsky}%

\[
\left(  \overline{\sum\epsilon_{i}}\right)  _{strutinsky}=\int\epsilon
\overline{g}_{M,\gamma}(\epsilon)d\epsilon
\]
Here, $\overline{g}_{M,\gamma}$ is the level density and $M$ and $\gamma$ are
respectively the so called order and smearing parameter of the Strutinsky's
procedure. The major defect of this method is that generally the results are
usually more or less dependent on these two parameters. A method to diminish
this dependence is to use the plateau condition, however in the case of finite
wells this is not systematically guaranteed. In this respect, it has been
demonstrated in ref. \cite{2006azizi} that the level density given by the
Strutinsky method is nothing but an approximation of the semiclassical level
density, i.e. a quantum level density from which the shell effects have been
washed out. Consequently, even though the Strutinsky is simpler in practice,
it is more interesting to work straightforwardly with the semiclassical
density because the problem of the dependence on the two above parameters is
in this way avoided. Thus, it is simply recommended to perform the smoothing
procedure with the semiclassical level density. The previous formula becomes
in this case:%

\[
\left(  \overline{\sum\epsilon_{i}}\right)  _{sc}=\int\epsilon g_{sc}%
(\epsilon)d\epsilon
\]
where $g_{sc}$ is the semiclassical level density.\newline Even though it is
not so obvious to derive a semiclassical density of states from a given
quantum Hamiltonian, there is for our case a rigorous solution (in the sense
where it the same quantum Hamiltonian which is "treated" semiclassically).
Indeed, for exactly the same Hamiltonian employed to determine the
eigenstates, the semiclassical level density is deduced following the
Wigner-Kirkwood method. The latter is based on the Thomas-Fermi approximation
plus a few corrections appearing as a power series of $(1/\hbar).$ In this
theory, the particle-number is expressed as a function of the Fermi level as
follows \cite{1975jenningsnp}:%
\begin{align}
N(\lambda)  &  =\frac{1}{3\pi^{2}}\left(  \frac{2M}{\hbar^{2}}\right)
^{3/2}\int_{D}d^{3}r\times\nonumber\\
&  \left\{  \left(  \lambda-V\right)  ^{3/2}+\frac{\hbar^{2}}{2M}\left[
\frac{3}{4}\kappa_{j}^{2}\left(  \nabla S\right)  ^{2}\left(  \lambda
-V\right)  ^{1/2}-\frac{1}{16}\left(  \lambda-V\right)  ^{-1/2}\nabla
^{2}V\right]  \right\}  \label{213}%
\end{align}
where $V$ and $S$ are the central field (including the coulomb potential for
the protons) and the spin-orbit field (see Ref. \cite{2004azizi}). The
classical turning points are defined by $\lambda-V(\overrightarrow{r_{sc}}%
)=0$. The domain of integration is defined by:
\begin{equation}
D:V(\overrightarrow{r})\leq\lambda=V(\overrightarrow{r_{sc}}) \label{domaine}%
\end{equation}
The semiclassical level density is thus derived as follows:%
\begin{equation}
g_{sc}(\epsilon)=\frac{dN(\epsilon)}{d\epsilon} \label{214}%
\end{equation}
\newline and the semiclassical energy is therefore:\newline%
\begin{equation}
\left(  \overline{\sum\epsilon_{i}}\right)  _{sc}=E_{sc}=%
{\displaystyle\int_{-\infty}^{\lambda}}
\epsilon g_{sc}(\epsilon)d\epsilon\label{215}%
\end{equation}
As already mentioned, the Fermi level is obtained from the following equation:%
\begin{equation}
N(\lambda)=N_{0} \label{nombre de p}%
\end{equation}
where $N_{0}$ is the particle-number (neutrons or protons).\newline The
semiclassical energy which is of course free from shell effects can be cast
under a power series of $(1/\hbar)$:%
\begin{align}
E_{sc}  &  =\lambda N_{0}-(E_{-3}^{0}+E_{-1}^{0}+E_{1}^{0})-(E_{-1}^{SO}%
+E_{1}^{SO})\label{216}\\
E_{-3}^{0}  &  =\frac{2}{15\pi^{2}}\left(  \frac{2M}{\hbar^{2}}\right)
^{3/2}\int_{D}d^{3}r\left(  \lambda-V\right)  ^{5/2}\label{217}\\
E_{-1}^{0}  &  =-\frac{2}{24\pi^{2}}\left(  \frac{2M}{\hbar^{2}}\right)
^{1/2}\int_{D}d^{3}r\left(  \lambda-V\right)  ^{1/2}\nabla^{2}V\label{218}\\
E_{-1}^{SO}  &  =\frac{2}{24\pi^{2}}\left(  \frac{2M}{\hbar^{2}}\right)
^{1/2}\int_{D}d^{3}r\left(  4\kappa_{j}^{2}\right)  \left(  \lambda-V\right)
^{3/2}\left(  \nabla S\right)  ^{2} \label{219}%
\end{align}
where for example $E_{-3}^{0}$ contains the term $\left(  1/\hbar\right)
^{3}$, etc... Here $SO$ means the term related to the spin-orbit
interaction.\newline The expressions of $E_{1}^{0}$ et $E_{1}^{SO}$ are very
complicated and become simple only for the non-deformed case (spherical
shape). The importance of these terms decreases rapidly. The r\'{e}f\'{e}rence
\cite{1975jenningsnp} gives the following percentages with respect to the
total semiclassical energy: $\lambda N_{0}\simeq30.5\%,$ $E_{-3}^{0}%
\simeq66\%,$ $E_{-1}^{0}\simeq2\%,$ $E_{-1}^{SO}\simeq1.25\%,$ $E_{1}%
^{0}\simeq0.05\%,$ $E_{1}^{SO}\simeq0.2\%$. In addition, it is to be noted
that the "active part" due to the deformation is even smaller. For this reason
the contributions $E_{1}^{0}$ et $E_{1}^{SO}$ (which are not given explicitly
here) are simply approached by their values for the spherical shape:%
\begin{align}
E_{1}^{0}  &  =\frac{1}{120\pi}\left(  \frac{\hbar^{2}}{2M}\right)  ^{1/2}%
\int_{D}drr^{2}\left(  \lambda-V\right)  ^{-1/2}\nonumber\\
&  \times{\Huge \{}\frac{11}{3}\frac{1}{r^{2}}\frac{d^{2}V}{dr^{2}}-\frac
{5}{12}\frac{d^{4}V}{dr^{4}}+\frac{7}{6}\frac{1}{r}\left(  \frac{d^{2}%
V}{dr^{2}}\right)  ^{2}\left(  \frac{dV}{dr}\right)  ^{-1}\nonumber\\
&  +\frac{7}{6}\frac{d^{2}V}{dr^{2}}\frac{d^{3}V}{dr^{3}}\left(  \frac{dV}%
{dr}\right)  ^{-1}-\frac{7}{12}\left(  \frac{d^{2}V}{dr^{2}}\right)
^{3}\left(  \frac{dV}{dr}\right)  ^{-2}{\Huge \}} \label{e1}%
\end{align}%
\begin{align}
E_{1}^{SO}  &  =-\frac{\kappa_{j}^{2}}{24\pi}\left(  \frac{\hbar^{2}}%
{2M}\right)  ^{1/2}\int_{D}drr^{2}\left(  \lambda-V\right)  ^{-1/2}\left(
\frac{dS}{dr}\right)  ^{2}\frac{d^{2}V}{dr^{2}}\nonumber\\
&  +\frac{1}{3\pi}\left(  \frac{\hbar^{2}}{2M}\right)  ^{1/2}\int_{D}%
drr^{2}\left(  \lambda-V\right)  ^{1/2}\nonumber\\
&  \times{\Huge \{}\kappa_{j}^{2}\left[  \frac{1}{2}\frac{dS}{dr}\frac{d^{3}%
S}{dr^{3}}-\frac{1}{r}\frac{dS}{dr}\frac{d^{2}S}{dr^{2}}-\frac{2}{r^{2}%
}\left(  \frac{dS}{dr}\right)  ^{2}\right] \nonumber\\
&  -\kappa_{j}^{3}\frac{1}{r}\left(  \frac{dS}{dr}\right)  ^{3}+\frac
{\kappa_{j}^{4}}{2}\left(  \frac{dS}{dr}\right)  ^{4}{\Huge \}} \label{kap}%
\end{align}
The different integrals (\ref{217}), (\ref{218}), (\ref{219}) are calculated
by the three dimensional Gauss-Legendre quadrature formulae. The set of
lattice points must verify Eq. (\ref{domaine}). In fact, for convenience, in
each direction, each interval is divided in elementary intervals in which the
quadrature formula is applied with a restricted number of nodes. The number of
points is increased in such a way to obtain stable numerical results.\newline
The Fermi level is not determined straightforwardly from Eq.
(\ref{nombre de p}), but solved as follows:\newline From:\newline$E_{sc}=%
{\displaystyle\int_{-\infty}^{\lambda}}
\epsilon g_{sc}(\epsilon)d\epsilon=%
{\displaystyle\int_{-\infty}^{\lambda}}
\epsilon\frac{dN(\epsilon)}{d\epsilon}d\epsilon$\newline a simple integration
by parts gives:\newline$E_{sc}(\lambda)=\lambda N(\lambda)-%
{\displaystyle\int_{-\infty}^{\lambda}}
N(\epsilon)d\epsilon$ with $g_{sc}(-\infty)=0$.\newline with the condition of
the Fermi level $N(\lambda)=N_{0}$, we will have\newline$E_{sc}(\lambda
)=\lambda N_{0}-%
{\displaystyle\int_{-\infty}^{\lambda}}
N(\epsilon)d\epsilon$\newline the differentiation with respect to $\lambda$
gives\newline$\frac{dE_{sc}(\lambda)}{d\lambda}=N_{0}-N(\lambda)=0$\newline
This means that for the constraint $N(\lambda)=N_{0}$, the value of $\lambda$
is the one which makes $E_{sc}(\lambda)$ minimum. Consequently, for a fixed
$N_{0}$ it is sufficient to look for this minimum with the help of Eq.
(\ref{216}) (this is what is done in the fortran program) without employing
subsidiary Eq.(\ref{nombre de p}). Knowing $\lambda$, the correctives terms
$E_{1}^{0}$ et $E_{1}^{SO}$ are deduced in the spherical approximation (as
mentioned before, the dependence on the deformation being very small for these
terms).\newline Unlike the previous case, the integral (\ref{e1}) and
(\ref{kap}) are one-dimensional and are also treated by Gauss-Legendre
formula. It is to be noted that the nodes of the quadrature do not make any
problem for the term $\left(  \lambda-V\right)  ^{-1/2}$, i.e., we have always
$\lambda\gg V(r_{node})$.

\section{Detailed expressions of $\left(  \nabla S\right)  ^{2}$ and
$\nabla^{2}V$}

Expressions $\left(  \nabla S\right)  ^{2}$ and $\nabla^{2}V$ are derived
analytically, for $\left(  \nabla S\right)  ^{2}$ the result is:%
\begin{align}
\left(  \nabla S\right)  ^{2}  &  =\dfrac{E^{2}}{(1+E)^{4}}\dfrac{1}%
{a_{so}^{2}}\left\{  \dfrac{x^{2}}{a^{4}}F(a)+\dfrac{y^{2}}{b^{4}}%
F(b)+\dfrac{z^{2}}{c^{4}}F(c)\right\} \label{221}\\
E  &  =\exp\left(  \dfrac{\sqrt{r_{2}}-1}{\sqrt{r_{4}}}\dfrac{\sqrt{r_{2}}%
}{a_{so}}\right)  ,r_{2}=\dfrac{x^{2}}{a^{2}}+\dfrac{y^{2}}{b^{2}}%
+\dfrac{z^{2}}{c^{2}},r_{4}=\dfrac{x^{2}}{a^{4}}+\dfrac{y^{2}}{b^{4}}%
+\dfrac{z^{2}}{c^{4}}\label{222}\\
F(\varepsilon)  &  =\left\{  \dfrac{1}{\sqrt{r_{4}}}-\dfrac{\left(
\sqrt{r_{2}}-1\right)  \sqrt{r_{2}}}{\left(  \sqrt{r_{4}}\right)
^{3}\varepsilon^{2}}+\dfrac{\sqrt{r_{2}}-1}{\sqrt{r_{4}}\sqrt{r_{2}}}\right\}
^{2} \label{223}%
\end{align}
with $S(\vec{r})=\kappa/\left[  1+\exp(R_{so}L_{so}/a_{so})\right]  .$ It is
worth to note that the spin-orbit coupling constant $\kappa$ (\cite{1973pauli}
and present work) is related to $\kappa_{j}$ of Ref. \cite{1975jenningsnp} by
the following equation:%
\begin{equation}
\kappa_{j}=\left(  -2Mc^{2}/\hbar^{2}c^{2}\right)  \kappa\simeq-0.0482\kappa
\label{les2kap}%
\end{equation}
This is due to the fact that in these references, the spin-orbit constant is
not defined in the same way. For $\nabla^{2}V$ we have:%
\begin{align}
\nabla^{2}V  &  =\left\{  \dfrac{2V_{0}E^{2}}{(1+E)^{3}}\dfrac{1}{a_{V}^{2}%
}-\dfrac{V_{0}E}{(1+E)^{2}}\dfrac{1}{a_{V}^{2}}\right\}  \left\{  \dfrac
{x^{2}}{a^{4}}F(a)+\dfrac{y^{2}}{b^{4}}F(b)+\dfrac{z^{2}}{c^{4}}F(c)\right\}
\nonumber\\
&  -\left\{  \dfrac{V_{0}E}{(1+E)^{2}}\dfrac{1}{a_{V}^{2}}\right\}  \left\{
D(x,a)+D(y,b)+D(z,c)\right\} \label{224}\\
D(\zeta,\varepsilon)  &  =\dfrac{1}{\varepsilon^{2}}F(\varepsilon
)+\dfrac{\zeta^{2}}{\varepsilon^{4}}G(\varepsilon)\label{225}\\
G(\varepsilon)  &  =-\dfrac{2}{\varepsilon^{2}\left(  \sqrt{r_{4}}\right)
^{3}}+\dfrac{3\left(  \sqrt{r_{2}}-1\right)  \sqrt{r_{2}}}{\varepsilon
^{4}\left(  \sqrt{r_{4}}\right)  ^{5}}-\dfrac{2\left(  \sqrt{r_{2}}-1\right)
}{\varepsilon^{2}\left(  \sqrt{r_{4}}\right)  ^{3}\sqrt{r_{2}}}\nonumber\\
&  +\dfrac{1}{\left(  \sqrt{r_{2}}\right)  ^{2}\sqrt{r_{4}}}-\dfrac
{\sqrt{r_{2}}-1}{\sqrt{r_{4}}\left(  \sqrt{r_{2}}\right)  ^{3}}\label{226}\\
V(\vec{r})  &  =\dfrac{V_{0}}{1+\exp(R_{V}\ L_{V}/a_{V})}(with\text{ }%
V_{0}<0)\label{227}\\
E  &  =\exp(R_{V}\ L_{V}/a_{V})
\end{align}
$F(\varepsilon)$ being defined by Eq. (\ref{223}). Finally, the shell
correction is calculated by replacing the Strutinsky's level density by the
semiclassical energy: This leads to:%
\begin{equation}
\delta E=%
{\textstyle\sum\limits_{i\text{ occup\'{e}s}}}
\epsilon_{i}-E_{sc} \label{correction couches}%
\end{equation}
The shell corrections are calculated separately for the neutrons and the
protons and then added to obtain the total shell correction.

\section{Pairing correction \label{bcs}}

We have took into account the pairing correction via the simple BCS
approximation. The Fermi level $\lambda$ and the gap parameter $\Delta$ are
solved from the well known system of coupled equations:%
\begin{equation}
\dfrac{2}{G}=\underset{k=1}{\overset{N_{P}}{\sum}}\frac{1}{\sqrt{\left(
\epsilon_{k}-\lambda\right)  ^{2}+\Delta^{2}}} \label{228}%
\end{equation}%
\begin{equation}
N\text{ or }Z=\underset{k=1}{\overset{N_{P}}{\sum}}\left(  1-\frac
{\epsilon_{k}-\lambda}{\sqrt{\left(  \epsilon_{k}-\lambda\right)  ^{2}%
+\Delta^{2}}}\right)  \label{229}%
\end{equation}
In these equations $G$ is the pairing strength and $\epsilon_{k}$ the
eigenvalues of the microscopic Hamiltonian. The upper index $N_{p}$ of the
sums represents the number of pairs of quasiparticles actually taken in the
calculations (with $N_{P}/2$\ above and below the Fermi level). $N_{P}/2$ is
the number of pairs of quasiparticles, taken in this work as the number of
levels between the Fermi and the first level of the spectrum.\newline For
convenience, we have adopted the prescription of Ref. \cite{1972bolsterli},
\cite{1972brack}, which has been widely used for realistic potentials such as
the Woods-Saxon potential (used here) or the folded-Yukawa potential
\cite{1972bolsterli} . In this prescription, the force of the pairing is
deduced from the empirical value of the gap $\overline{\Delta}=12/\sqrt{A}$
and from $N_{p}$ (see text just above):%
\begin{equation}
\frac{1}{G}\approx\overline{g}(\lambda)\ln\left(  \frac{N_{p}}{\overline
{g}(\lambda)\overline{\Delta}}\right)  \label{230}%
\end{equation}
Here $\overline{g}(\lambda)$ denotes the smoothed level density determined
from the Strutinsky's procedure or by a semiclassical method as in the present
work. The nonlinear system is solved by successive iterations until a given
precision. At each iteration, we deduce the occupation probabilities from new
couple $\lambda$ and $\Delta$ :%
\begin{equation}
\upsilon_{k}^{2}=\frac{1}{2}\left(  1-\frac{\epsilon_{k}-\lambda}%
{\sqrt{\left(  \epsilon_{k}-\lambda\right)  ^{2}+\Delta^{2}}}\right)  \text{,
\ \ \ \ \ \ \ \ \ }k=1,.........N_{p} \label{231}%
\end{equation}
conversely, from the "new" occupations amplitudes $\left(  u_{k},\upsilon
_{k}\right)  $ we deduce the "new" gap:%
\begin{equation}
\Delta=G\underset{k=1}{\overset{N_{P}}{\sum}}u_{k}\upsilon_{k}\text{
\ \ \ avec \ \ \ \ \ \ }u_{k}^{2}=1-\upsilon_{k}^{2} \label{232}%
\end{equation}
and so on\newline For one kind of particles, the pairing correction to the
liquid drop model is defined as:%

\begin{equation}
\delta P_{pairing-correct.}(N\text{ or }Z,\beta,\gamma)=P-\overline{P}
\label{233}%
\end{equation}
were%

\begin{equation}
P=\underset{k=1}{\overset{N_{P}}{\sum}}2\upsilon_{k}^{2}\epsilon_{k}%
-\frac{\Delta^{2}}{G}-\overset{N_{p}/2}{\underset{k=1}{\sum}}2\epsilon_{k}%
\end{equation}
is the usual energie for a correlated system of fermions, and%

\begin{equation}
\overline{P}=-\frac{1}{2}g_{sc}(\lambda)\overline{\Delta}^{2}%
\end{equation}
is its smooth part (i.e., without shell effects) assumed already contained in
the liquid drop model \cite{1972bolsterli}.

Finally, with obvious notation, the potential energy of deformation can be
summarized as follows:%

\begin{align*}
E_{def}(N,Z;\beta,\gamma)  &  =E_{LD}(N,Z;\beta,\gamma)+\delta E_{sc}%
(N;\beta,\gamma)+\delta E_{sc}(Z;\beta,\gamma)\\
&  +\delta P(N;\beta,\gamma)+\delta P(Z;\beta,\gamma)
\end{align*}

where the shell and pairing corrections are due to separates contributions of
neutrons and protons.

\section{Handling and numerical checking of the associated fortran code}

\subsection{The non-deformed (spherical) case}

Two codes have been built for calculating the semiclassical energy. The first
code is based on the general deformed case which consists of three fold
integral (subroutine scdefor) and the second can only be used for the
spherical shape with a one dimensional integral (subroutine sclspher1). Then,
it is possible to make a cross checking in the spherical (non-deformed) case.
To make further comparisons with other works, we have chosen the same examples
as those of the Ref. \cite{1975jenningsnp} . The different contributions to
the semiclassical energy Eq.(\ref{216}) are detailed in the following tables:

\begin{table}[ptbh]
\caption{Comparisons between our codes and other papers for the non-deformed
case. The energies are given in MeV.}%
\label{t1}
\begin{tabular}
[c]{l|l|l|l|l|l|l|l|l}\hline
& \multicolumn{8}{|l}{$N=126;$ $\ Vo=44MeV;$ $\ R_{V}=R_{so}=7.52fm;$
$\ a_{V}=a_{so}=0.67fm;$ $\ \varkappa_{j}=-0.7491;$ \ $Z=arbitrary$}\\\hline
routine & $E_{sc}$ & $\lambda$ & $\lambda N_{0}$ & $E_{-3}^{0}$ & $E_{-1}^{0}$
& $E_{-1}^{SO}$ & $E_{1}^{0}$ & $E_{1}^{SO}$\\\hline
scdefor (present code) & $-2282.56$ & $-5.7476$ & $-724.21$ & $1590.51$ &
$-50.90$ & $24.08$ & from sclspher1 & from sclspher1\\
sclspher1 (present code) & $-2282.61$ & from scdefor & $\ 724.21$ & $1590.51$
& $-50.87$ & $24.08$ & $-1.07$ & $-4.27$\\
Ref. \cite{1975jenningsnp} & $-2282.5$ & $-5.7323$ & $-722.27$ & $1592.42$ &
$-50.87$ & $23.92$ & $-1.07$ & $-4.24$%
\end{tabular}
\end{table}

\begin{table}[ptbh]
\caption{Analog calculations as in the table \ref{t1} for another example}%
\label{t2}%
\begin{tabular}
[c]{l|l|l|l|l|l|l|l|l}\hline
& \multicolumn{8}{|l}{$N=184;$ $\ Vo=43MeV;$ $\ R_{V}=R_{so}=8.48fm;$
$\ a_{V}=a_{so}=0.67fm;$ $\ \varkappa_{j}=-0.7321;$ \ $Z=arbitrary$}\\\hline
routine & $E_{sc}$ & $\lambda$ & $\lambda N_{0}$ & $E_{-3}^{0}$ & $E_{-1}^{0}$
& $E_{-1}^{SO}$ & $E_{1}^{0}$ & $E_{1}^{SO}$\\\hline
scdefor (present code) & $-3228.76$ & $-4.9328$ & $-907.63$ & $2359.67$ &
$-61.27$ & $30.12$ & from sclspher1 & from sclspher1\\
sclspher1 (present code) & $-3228.78$ & from scdefor & $-907.63$ & $2359.67$ &
$-61.26$ & $30.12$ & $-2.62$ & $-4.76$\\
Ref. \cite{1975jenningsnp} & $-3230.0$ & $-4.9281$ & $-906.77$ & $2360.52$ &
$-61.24$ & $30.13$ & $-1.47$ & $-4.97$%
\end{tabular}
$\ $\end{table}

The numerical values of the parameters of the potential are displayed in the
tables themselves. These calculations are performed for neutrons. The
dependence on the proton number appears only through the parameters of the
woods-saxon potential. Appart from numerical uncertainties due to different
numerical approaches, the results are found very close.

\subsection{The deformed case}

\begin{figure}[ptbh]
\includegraphics[width=160mm,keepaspectratio]{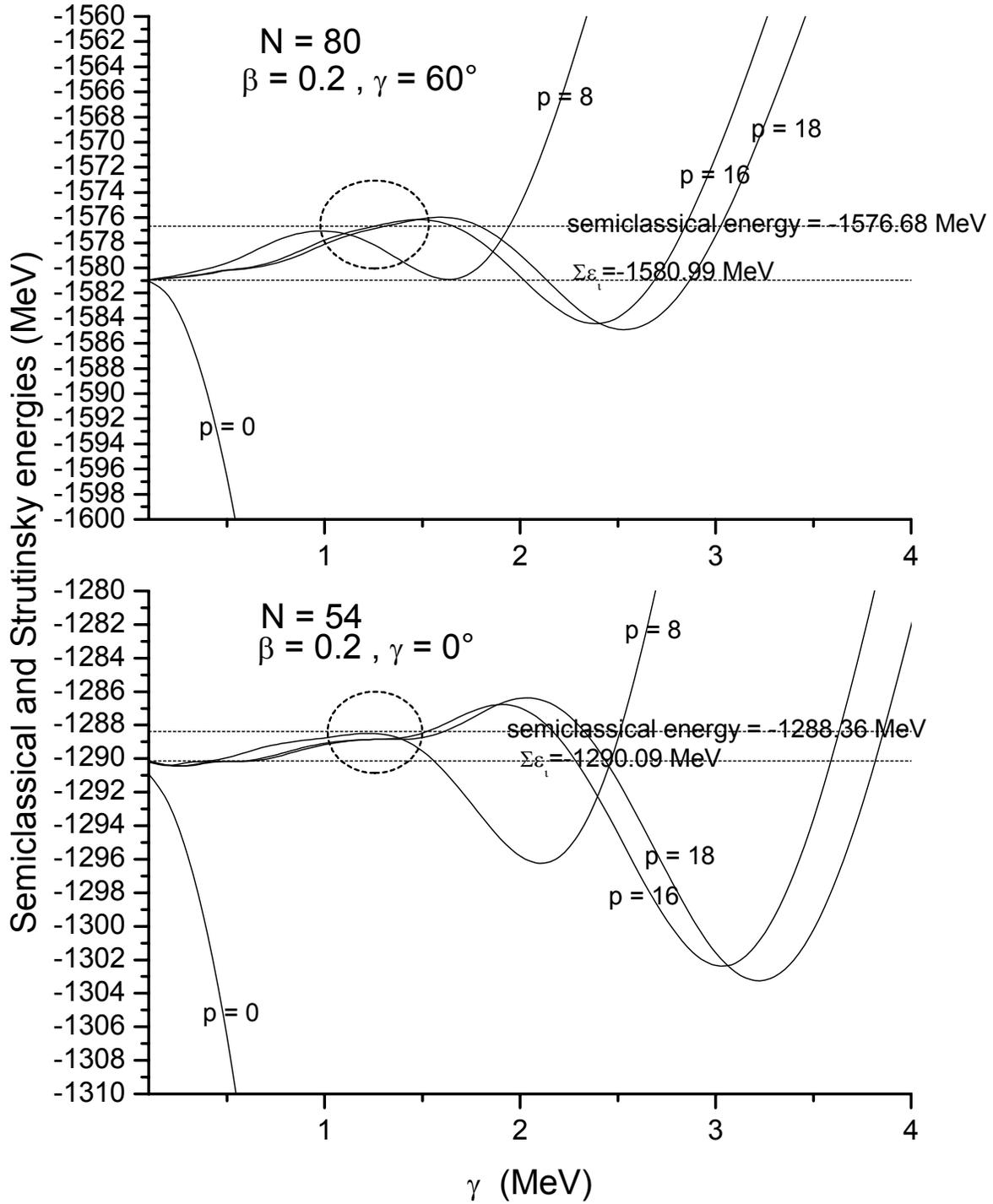}\caption{(Bottom)
Semiclassical and Strutinsky calculations of the energy as function of the
smearing parameter $\gamma$ for different orders of the curvature correction.
The semiclassical energy as well as the sum of single-particle energies are
given by a straight line. The calculations are made for N = 54 (Bottom) and
N=80 (Top).}%
\label{fig1}%
\end{figure}

To our knowledge, semiclassical calculations for the Hamiltonian such as the
one considered in this paper do not exist in the literature. For this reason
the only way to test the code in the deformed case is to compare the results
with those of the Strutinsky type. However, it is well known that the latter
method often gives results with some uncertainty. Consequently, as
demonstrated in Ref.\cite{2006azizi}, in performing these tests, we must keep
in mind that the Strutinsky calculations are only approximation of the
semiclassical limit. In this respect, the smallness of the relative error
gives a good idea on the quality of the results The essential point is to
verify that the code runs properly. In fact the code has been checked
extensively a longtime ago. As examples, we give two deformed cases in fig.
(\ref{fig1}). The parameters are given in the readme4.pdf file.

It is very clear that an approximative plateau exists in the region
$\hbar\omega\lesssim\gamma\lesssim1.5\hbar\omega$ represented by a circle. For
the order $p=0$ we do not obtain any plateau. In the region of the plateau the
relative error is less than $1MeV$ per $1300\sim1500MeV$ in the both cases.

\section{Data, input and output of the code}

This program has been designed on the Compac Visual Fortran version 6.6.0
(optimized settings). In fact, the structure of the code is somewhat
complicated. So, it is no need to give too much details. The essential point
is to handle the basic input data and to be able to read the desired data from
the output files. The fortran source code denoted by "enerdef.f" can be
downloaded from:
\textbf{http://macle.voila.fr/index.php?m=c9ae77e8\&a=7d397569\&share=LNK80764b6393d92f388}%

\subsection{Input data}

\subsubsection{Parameters of the Woods-Saxon potential (file
WS\_parameters.dat)}

The microscopic model and the associated FORTRAN code is the same as the one
of Ref.\cite{2004azizi} and \cite{2007azizi}. Therefore the parameters of the
woods saxon potential are read from the file parameters.dat. renamed in the
present work as ws\_parameters.dat.

See pdf file readme1\_woods saxon parameters

\subsubsection{Other input data (Beginning of the main program)}

See pdf file readme2\_input data.

They must be pr\'{e}cised at the beginning of the main program:

nmax=10 to 20 \ is linked to the size of the oscillator basis

iuno=1 (for single deformation) or 0 (for lattice mesh points)

if iuno=0 the three following data must be pr\'{e}cised:

betamax=0.0 up to about 1.0 is the maximal value of the parameter beta

ibetapoints= is the number of points (minus one) in the beta direction

igamapoints=is the number of points (minus one) in the gamma direction

\subsubsection{"Manual" input data (Keyboard)}

The kind of nucleons, the number of protons and the number of neutrons have to
be entered manually on the keyboard.

If iuno=1 the deformation must also be pr\'{e}cised in the terminal (do not
forget that the deformation parameters are real quantities)

\subsubsection{Liquid drop data (Module liquid drop)}

The fissility parameter and other miscellaneous data for the liquid drop model
are fixed in the subroutines eld, bbs,bbc in the module "liquid drop".

\subsubsection{Strutinsky calculations}

Additionally, this code is able to perform Strutinsky calculations.Two
routines are devoted to this task. The first (Nstrutinsky) solves the Fermi
level. The second (Strutinsky) calculates the smooth energy once the fermi
level is known from the first routine. The essential points are the following
for the rwo routines -(see readme3\_strut.pdf file):

ggam (input) = is the smearing parameter ($ggam=\gamma=41.A^{-1/3}MeV$)

the numbers 0,8,16,18 (input, up to 18) = correspond to the curvature
correction of the shell correction= does not exceed 18 (here four calculations
are done).

rnumb0, rnumb8, etc...(output for Nstrutinsky)= number of particle found after
solving equation=checking

hnew0,hnew8,....(output for Nstrutinsky and input for Strutinsky)= Fermil
level for different orders of the curv. correct.

res0,res8,....= shell correction for different orders of the curv. correct.

The code performs shell corrections in loop do for several values of ggam and
four values of the order of the curvature correction.

\subsection{Data checkings}

In addition, the input and output data for the checkings are detailed in the
readme4.pdf. file.

\section{Output data}

The files eigenvalues and eigenvectors give the solutions of the Schrodinger
equation. All results are given separately for neutrons and protons. Due to
the coulomb interaction, the calculations in the proton case are significantly
slower. However for a family of isotope the calculations for the protons must
be taken only once.

The files:

del\_n.dat, del\_p.dat

eldm\_n.dat, eldm\_p.da

epot\_n.dat, epot\_p.dat

give in the third column respectively the gap parameter, the energy of the
liquide drop model and the deformation energy (all in MeV) for neutrons (\_n)
and protons (\_p). The two first columns specify the deformation in the
sextan,beta-gamma. Gamma is given in degrees.

The files control results\_n.dat and control results\_p.dat give some details
of the calculations.

The File 2000n.dat (neutrons) or 2000p.dat (protons) gives the shell
correction (columns 2 to 5). Each column corrresponds to a given order of the
shell correction. Each row corresponds to a given value of the smearing
parameter. The first column gives the smearing parameter (in hW units) and the
last column gives the semiclassical value of the energy.

\end{document}